\documentstyle[12pt]{article} %preprint
\input{epsf}

\begin{document}

%\let\nopictures=Y

%\draft

%\preprint

%\nopagebreak
\title {ASYMPTOTIC SERIES AND PRECOCIOUS SCALING}

\author{
Geoffrey B. West \\
%\footnote{To whom correspondence should be addressed. email: gbw@lanl.gov.},  
%James H. Brown$^{2,3}$, 
%Brian J. Enquist$^{2,3}$
 %\vspace{.125in} 
 Theoretical Division, MS B285, \\
Los Alamos National Laboratory, \\
Los Alamos, NM  87545, USA.\\
%$^2$ The Santa Fe Institute,
%1399 Hyde Park Road,
%Santa Fe, NM 87501, USA.\\
%\vspace{-.125in}
%$^3$ Dept. of Biology,
%University of New Mexico,
%Albuquerque, NM  87131, USA. \\
}

\vspace{5.0cm}

\maketitle
\vspace{.125in}

%\pagebreak

\def\refname{REFERENCES AND NOTES}

\vspace{1.0cm}

\begin{abstract}

Some of the basic concepts regarding asymptotic series are reviewed. A heuristic proof  is
given that the divergent QCD perturbation series is asymptotic. By treating it as an
asymptotic expansion we show that it makes  sense to keep only the first few terms. The
example of
$e^+e^-$ annihilation is considered. It is shown that by keeping only the first few terms
one can get within a per cent (or smaller) of the complete sum  of the series even at very
low momenta where the coupling is large. More generally, this affords an explanation of the
phenomena of precocious scaling and  why keeping  only leading order corrections generally
works so well.

\end{abstract}

\vspace{.5in}

%\narrowtext
%\widetext
%\twocolumn
\pagebreak

By virtue of its property of asymptotic freedom QCD perturbation theory gives an excellent
description of many high energy processes involving large momentum transfers, especially in
inclusive phenomena. The agreement with theory often remains valid down to surprisingly low
momenta as in the classic case of precocious scaling in deep inelastic lepton scattering
where approximate scaling persists down to less than a GeV. At first sight the great
success of perturbation theory is all the more surprising given its relatively large
coupling constant (g) and the fact that the series is divergent. Furthermore, it is
generally believed that the nature of the divergence is sufficiently severe that the series
is not summable by conventional methods: no available technique exists
for reconstructing a unique analytic representation for its sum. This is in marked
contrast to scalar field theories which have been shown to be Borel summable and for
which a well-defined representation exists \cite{sum}. Thus a major theoretical question
concerning  QCD  is how to control and make sense out of the
non-summable divergence in its perturbation expansion. Ultimately  large order loop
contributions must dominate the leading terms so  a well-defined procedure justifying the
use of just the leading order estimates is required. Equally as important is to understand
the errors  incurred by such a procedure. A natural framework for dealing with this problem
is to treat the QCD perturbation series as  asymptotic  and apply an Euler-Poincar{\' e} 
analysis to estimate the number of terms that should be kept in order to obtain an optimal
estimate of its sum\cite{asymp}. A by-product of such an analysis is an estimate of the
error. In what follows we first  give a heuristic proof that the series is indeed
asymptotic. Using generic forms for estimates of the large order coefficients optimal sums
for the series are subsequently obtained. It is worth  emphasizing that the presumption of an
asymptotic series is significantly weaker than assumptions required for summability.

                Generally speaking, the nature of an asymptotic series is such that, as  the
first $N$ terms are calculated, the correct sum is uniformly approached until, after an
optimum number of terms, $N_0$, is reached,  adding additional terms drives the sum
further and further from the correct result ultimately leading to a divergence. A
crucial characteristic of  this behaviour is that by retaining only the optimal $N_0$ terms
and discarding the rest, one can get exponentially close to the  correct sum when the
expansion parameter is small. We shall show that carrying through such an analysis for QCD
leads to the following conclusions:

\begin{itemize}

\item[i)]{ $N_0$ is a relatively small number which depends on the
characteristic momentum.}

\item[ii)]{At high energies, where $\alpha_s (\equiv g^2/4\pi)\sim 0.1$, the error incurred
by keeping only
$N_0$ terms is much less than a per cent.}

\item[iii)]{At lower energies, where $\alpha_s \sim 0.2$, say, $N_0$ decreases and can
become less than 2; the corresponding error can still be less than a per cent.}

\end{itemize}

Thus, at low energies where  $\alpha_s$ is quite large, the sum of the perturbative series
can be approximated to within a per cent by keeping only the first couple of terms! Adding
higher order loop contributions will only drive one further from the correct
result. Since the error is only  roughly a per cent or so, it is still  well
within      typical experimental errors. This, therefore, offers a possible
explanation as to why perturbation theory works so well even at rather modest energies and
momenta where $\alpha_s$ is not so small. The nature of these conclusions is rather
general (when applied to appropriate processes); the details, however, will be process
dependent and, in general, depend on the effective expansion parameter.

   Before deriving these results we first review the definition and some salient properties
of asymptotic series. Consider a function $F(z)$       which is analytic
in a wedge centered on the origin. If the wedge opening is less than $2\pi$, then $F(z)$ is 
non-analytic at the origin and so a power series development

\begin{equation}
F(z) \approx \sum^{\infty}_0 a_nz^n
\label{one}
\end{equation}
must have zero radius of convergence. For QCD perturbation theory, $z\equiv g^2$ and the
$a_n$ are the sum of all Feynman graphs at order $n$. For fixed $z$ the remainder 

\begin{equation}
R_N(z) \equiv |F(z) - \sum^{N}_0 a_nz^n|
\label{two}
\end{equation}
diverges for large $N$. On the other hand, if the series is asymptotic, then, for a given
$N$, $R_N(z)/z^N \sim 0$ when $z\sim 0$. This follows from the basic definition of an
asymptotic series\cite{asymp}, namely that, for a fixed $N$ and  $z\sim 0$,

\begin{equation}
R_N(z) \leq C_N z^{N+1} 
\label{three}
\end{equation}
Now let us exploit, the analytic structure of $F(z)$ by deriving a generalized dispersion
relation. Consider some point $z$ encircled by a contour $C$ in a
region where $F(z)$ is everywhere analytic and write a standard Cauchy representation

\begin{equation}
F(z) = \int_C {dz'\over{2\pi i}} {F(z')\over {z' - z}}  
\label{four}
\end{equation}
The contour can be distorted to encircle all the singularities of $F(z)$ leading to

\begin{equation}
F(z) = \int_L {dz'\over{2\pi i}} {f(z')\over {z' - z}}  
\label{five}
\end{equation}
where $f(z) \equiv (2\pi i)^{-1}\sum$disc $F(z)$ with the sum and integral being taken
over all discontnuities of $F$. Possible contributions from the contour at infinity have
been dropped; if the integrals are not sufficiently convergent for this to be valid,
        sufficient subtractions are assumed to have been made. Their presence does not
 change the general argument. Formally expanding (\ref{five}) in powers of $z$
leads to

\begin{equation}
a_n = \int_L {dz\over z^{n+1}} f(z)
\label{six}
\end{equation}
This is the basic formula used to derive the $a_n$. Indeed, if a path integral
representation is used for $F(z)$ then this formula reproduces conventional
Feynman graphs. It can also be used to estimate the large n behavior of $a_n$ via a saddle
point technique. The path integral representation for $F(z)$ has the generic
structure  $\int D\phi e^{-S[\phi]/z}$, where $S$ is the action; [here the generic
field $\phi$  stands for all fields including fermions and gauge bosons]. Thus, singularities
in  $z$  develop only
when Re $z < 0$, i.e., along the negative real axis in which case $f(z)$ reduces to
$2i$ Im$F(z)$\cite{lipatov}. This relatively simple structure is broken by renormalization
which gives rise to a considerably more complex singularity structure\cite{sum}. This
situation makes it difficult to give a rigorous proof that the series must be asymptotic;
nevertheless, a heuristic proof can be given.

                The crucial observation is that Eqs. (\ref{two}) and (\ref{five}) can be
combined to give 

\begin{equation}
 R_N(z) = z^{n+1}\int_L {dz' f(z')\over {z'^{n+1}(z' - z)}}
\label{seven}
\end{equation}
If, when $z \sim 0$, the integral exists, then the constraint of Eq.(\ref{three})
is clearly satisfied; however, in this limit the integral is, at least formally, simply
$a_{N+ 1}$ [see Eq. (\ref{six})]. Furthermore, these coefficients, $a_{n}$, must exist,
since they represent the sum of all Feynman graphs of a given order, and can, in fact, be
generated from Eq. (\ref{six}). Thus, as $z \sim 0$ 

\begin{equation}
R_N (z) \sim a_{N+ I} z^ {N+1}
\label{eight}
\end{equation}
which manifestly satisfies (\ref{three}) thereby showing that the series must be
asymptotic \cite{foot1}. Q.E.D.

The expression (\ref{eight}) manifests one of the essential features of asymptotic series:
as $N$ increases, $a_N$ grows whereas $z^N$ diminishes, thereby leading to a minimum value
$R_N$. Before applying this to QCD let us first briefly review the relationship of these
ideas to summability.

The idea behind Borel summability is that, although the series (\ref{one}) may be
divergent, the series
\begin{equation}
G(z) \equiv \sum^{\infty}_0 {a_n\over {n!}}z^n
\label{nine}
\end{equation}
may be convergent. In that case $F(z)$ can formally be obtained from G(z) by the
Laplace transform

\begin{equation}
F(z) ={1\over z}\int^{\infty}_0due^{-u/z}G(u)
\label{ten}
\end{equation}
This then serves to define  a function that is asymptotic to the original divergent series.
Under certain restrictive conditions (loosely speaking, that the integral and its transform
exist) this construction defines a unique function.

As already mentioned it is believed that the Borel procedure can be applied
to scalar field theories to give a representation for its perturbative sum; such
is not the case, however, for QCD. The difference between the two can be
characterized by the c1assic examples (i) $a_n =(-1)^n n!$ for which $G(z)=(1 + z)^{-1}$ and
(ii)  $a_n = n!$ for which $G(z)  =  (1 - z)^{-1}$. The latter, which is the analog of QCD,
is not Borel summmable since the integral is not well-defined because of the ambiguity
in how to treat the singularity at $u = 1$\cite{foot2}. The former, however, which is the
analog to scalar field theory, is well-defined. From Eq. (\ref{ten}) we deduce that

\begin{equation}
E(z) =\int^{\infty}_0{dve^{-v}\over{1+vz}}
\label{eleven}
\end{equation}
is the analytic reconstruction of the divergent series

\begin{equation}
1 - z + 2!z^2 - 3! z^3 + 4!z^ 4 - ......
\label{twelve}
\end{equation}
Eq. (\ref{eleven}) can therefore be used to define and evaluate the divergent sum, Eq.
(\ref{twelve}). Thus, for example, a numerical evaluation gives $E(0.1) = 0.915633$.......
However, and this is the important point, it is also possible and, in general, considerably
easier and more efficient to use the series directly to get an excellent estimate for
$E(z)$. To see how this comes about, we return to  Eq. (\ref{eight}) from which we see that,
for small $z$, $R_N(z)\approx (N+1)!z^{N+1}$. Thus $R_N$ at first decreases  as
$N$ increases, reaching a minimum at $N=N_0$, where $\partial R_N(z)/ \partial N =0$, 
and then increases in an unbounded fashion. The minimum occurs at $N = N_0\approx
(1/z)e^{1/2z} - 2$ at which point
\begin{equation}
R_{N_0} \approx [2\pi e(N_0 + 2)]^{1/2}e^{-(N_0 + 2)} \approx (2\pi/z)^{1/2}e^{-1/z}
\label{thirteen}
\end{equation}
This is a remarkable result, for it shows that one can get exponentially close to
the correct answer by keeping only the first $N_0$ terms. Eq. (\ref{thirteen})
thus represents the closest one can get to the correct answer by this technique.
So, if   $z =  0.1$, as in the above  example, this gives $N_0\approx 8.5$. Keeping 8 terms
in (\ref{twelve}) then gives $E(0.1)\approx 0.915819$ whilst 9 terms gives $E(0.1)\approx
0.915460$ in excellent agreement with the exact result obtained by numerically evaluating
the integral. The error thus incurred is in agreement with Eq. (\ref{thirteen}) which gives 
$R_{N_0}\approx 3.6\times 10^{-4}$. Now, suppose that $z= 0.2$, then $N_0\approx 5.5$ so only
3 - 4 terms  need be kept!
The error thus generated is thus only $R_{N_0}(0.2) \approx 3.78 \times 10^{-2}$. Keeping 3
or 4 terms one trivially finds  that, $E(0.2)$
lies between 0.832 and 0.870 to be compared with the exact, number 0.8521 ... Thus, if 
a given
accuracy is sufficient one need only keep a relatively small number of terms in the
series. Even when $z\approx 0.2$ one can get within
a few per cent of the exact answer.  In such a situation, there is no advantage
in having an  explicit
representation such as (\ref{eleven}) and little or  nothing would be  lost without it. 
We now apply these ideas to QCD and show how the asymptotic nature of the series
and the large value of $\alpha_s$
can be put to advantage.

                As an illustrative example consider the QCD
contribution to the total $e^+e^-$  annihilation
cross-section\cite{west}. This is usually expressed as a ratio, $R$, normalized to the
cross-section for  $\mu^+\mu^-$  production. Asymptotic freedom dictates that its high
energy behaviour is ultimately given by $\sum Q_i^2$, $Q_i$ being the  charge of
the $i$th quark species. Leading corrections to this can be developed in a standard
perturbative series:

\begin{equation}
R\bigg[{q^2\over \mu^2},\alpha_s(\mu)\bigg]\approx \bigg(\sum Q_i^2\bigg)\sum^\infty_0
r_n\bigg({q^2\over\mu^2}\bigg) \bigg({\alpha_s\over\pi}\bigg)^n
\label{fourteen}
\end{equation}
Here $q$ is the four-momentum delivered by the $e^+e^-$ pair; we have also made  manifest
the arbitrary scale $\mu$ needed to define  $\alpha_s$.         
The first four coefficients of this series have been calculated: when $q^2 = \mu^2$      
they are $r_0 = r_1 = 1;        r_2  = 1.41$ and
$r_3 = -12.8$. The first two are both scheme and momentum independent whereas
the numbers quoted for $r_2$ and $r_3$ are in the modified-minimal-subtraction ($\bar {MS}$)
scheme with five flavors.

It is generally agreed \cite{sum} that for large $n$ the structure of the $r_n$ has the
generic form
 
\begin{equation}
r_n(1) \approx Ca^nn^b\Gamma(n+c)
\label{fifteen}
\end{equation}
 where $C, a, b$ and $c$ are constants.
This can be motivated from the path integral representation where all of the coupling
constant dependence resides in the factor $e^{-S/g^2}$. Note that this implies that the real
effective expansion parameter is $a\alpha_s/\pi$ rather than $\alpha_s/\pi$. From
 Eq. (\ref{fifteen}) we can also estimate the optimum number of terms to be
kept in order to get as close as possible to the exact sum:

\begin{equation}
N_o \approx a^{-1} -1 -c         
\label{sixteen}
\end{equation}
as well as the corresponding error:

\begin{equation}
R_{N_o} \approx (2\pi)^{1/2}Ca^{1/2-b-c}e^{-1/a}
\label{seventeen}
\end{equation}

Now, renormalisation, which introduces the arbitrary  scale
$\mu$ into the problem, forces
$R$ to be a function of  the single variable
$z\equiv({q^2/\mu^2})e^{-\int dg/\beta (g)}$ rather than of the two  variables
${q^2/\mu^2}$ and 
$\alpha_s(\mu)$ separately. Here $\beta (g)$ is the usual $\beta$-function which has the
perturbative expansion: $\beta (g) \approx -g^3(b_1 + b_2 g^2 + \dots)$. Thus $z\approx
({q^2/\mu^2}) e^{1/{b_1 g^2}} {g^{b'}}$ where $b'\equiv b_2/b_1^2$. This  suggests that the
real expansion parameter is not $\alpha_s/\pi$ but
rather ${b_1 g^2} = 4\pi b_1\alpha_s$ since these must always occur together. We have argued
\cite{west} from a more detailed analysis based on 
$q^2$ analyticity and the use of a saddle point technique  that
it is, in fact,
$4\pi^2 eb_1\alpha_s$  so that, in terms of Eq. (\ref{fifteen}), $a = 4\pi^2 eb_1$, 
$b = -2$ and $c = b'$.  The general structure of the
results and  conclusions presented here  do not depend on the detailed values of the
coefficients. Although there has been some
criticism\cite{comment} of the derivation of the estimates in \cite{west} it is the generic
structure represented by Eq. (\ref{fifteen}) with $a\sim 4\pi^2 b_1$ that drives the
general conclusions. Note also that an expansion around instanton solutions, where
$S\rightarrow {8\pi^2}$ would suggest a value of $a =1/2$ which is significantly smaller
than the value derived from the above  renormalisation argument. These latter
``non-perturbative'' effects (``renormalons'') can therefore be  expected to dominate the
large
$n$ behaviour arising from instantons.

We can therefore estimate that $N_o\sim (4\pi^2 eb_1\alpha_s)^{-1} - b' -1$. With $\alpha_s
\sim 0. 15$, this gives $N_o \approx 5$. Thus it makes sense  to keep only 5 terms at best
in the series. Keeping more than this, even if  calculable, drives one further from
the correct sum. 
An estimate\cite{west} of $C$ which gives good agreement with $r_3(1)$ is $(-4\pi^3 b_1
e^{-b'})^{-1}$ With this the miminum error is found to be $ 7 \times 10^{-4}$, i.e. less
than  $0.1\%$. This is illustrated in Table I and Fig.l. 

The discussion so far has focused on the single value  $q^2 = \mu^2$. Indeed, all of the
numbers quoted are scale dependent, the scale being determined by the value of $\mu$ for
which $\alpha_s = 0.15$. It is conventional to use the invariance of R to changes in $\mu$
to transfer the $q^ 2$ dependence from the $a_n$ to $\alpha_s$ by introducing a running
coupling constant:

\begin{equation}
R\bigg[{q^2\over \mu^2},\alpha_s(\mu)\bigg] = R\bigg[1,\bar\alpha_s(q^2)\bigg] 
=\bigg(\sum Q_i^2\bigg)\sum^\infty_0 r_n(1)\bigg[{\bar\alpha_s(q^2)\over\pi}\bigg]^n
\label{eighteen}
\end{equation}
 where $\int^{\bar\alpha_s}_{\alpha_s}d{\alpha_s}/\bar\beta(\alpha_s) = \ln (q^2/\mu^2)$
with $\bar\beta(\alpha_s)\equiv g\beta(g)/16\pi^2$.  This is usually expressed in terms of
a QCD scale parameter $\Lambda$: $\alpha_s(q^2) = (4\pi b_1 t)^{-1}[1-b'\ln t/t +\cdots]$,
where $t\equiv \ln(q^2/\Lambda^2)$.
The analysis that was applied to the original series in terms of $\alpha_s(\mu)$ can now
be applied to Eq. (\ref{eighteen}). The only difference is that $\alpha_s$ is replaced by
$\bar\alpha_s(q^2)$ so that the corresponding $N_o$ now becomes momentum dependent. From
Eq. (\ref{sixteen}) we      obtain an expression for the $q^ 2$ -dependence of the optimum
number of terms for the series  (\ref{eighteen}):

\begin{equation}
\bar N_o \approx e^{-1}(t + b'\ln t) +1/2 -b' 
\label{nineteen}
\end{equation}
For simplicity we have  kept here only the leading term; corrections have only a small
effect. The corresponding error at this optimum number is given approximately by 
$\bar R_{N_o}(q^2)\approx (8\pi^5)^{-1/2}b_1^{-1}(\Lambda^2/q^2)^{1/e}$. 
These equations exhibit the behavior already alluded to, namely that,
as $q^2$ decreases and $\bar\alpha_s(q^2)$ increases, the optimum number
of terms that can be retained actually decreases. The price to be paid for this remarkable
behaviour is that the error thus incurred correspondingly increases.

\begin{figure}
\begin{center}
\epsfxsize=0.9\hsize\epsfbox{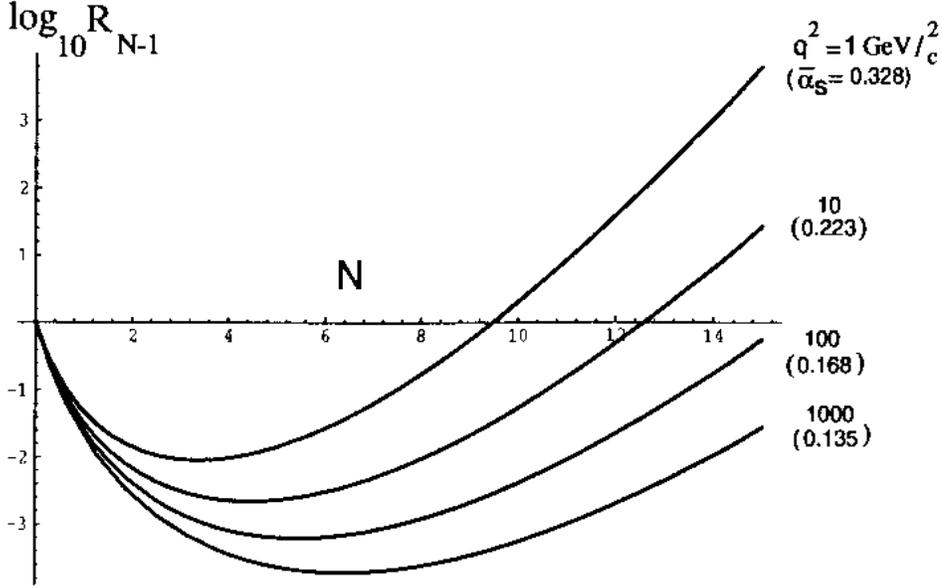}
\end{center}

% \epsfbox{west.curves.150.eps}

\caption{Plot of log$_{10}$ $R_{N-1}$ vs. $N$ for various values of $q^2$ with $\Lambda =
100$MeV. The curves are labeled by both the values of $q^2$ in GeV/$c^2$ and the equivalent
value of $\bar\alpha_s$. Notice that the error in keeping only the one-loop correction
($N=2$) is always $\leq 1\%$ even at very low values of $q^2$. The locus and minimum value
are approximately given in the text.}
\end{figure}

This pattern is shown in Fig. 1 where the error $\bar R_{N}$ is plotted
versus $N$ for various values of $q^2$ with $\Lambda =100$MeV. As can
readily be seen from the graph only a very few terms need be retalned in order to get
within a few per cent of the correct sum even down to very low values of $q^ 2$.
Explicit non-perturbative corrections to the sum arising, for example, from instantons
are expected to be of order $e^{-2\pi/\alpha_s}$ which is much smaller than even the
smallest value of $R_N$. This call be put slightly differently by noting that such
instanton contributions behave Ilke $(\Lambda^2/q ^2)^{-8\pi^2 b_1}$, which remains
much smaller than $\bar R_{N_o}(q^2)$ down to values of $q^2\sim  1$GeV/$c^2$. This
therefore serves as a possible explanation as to why tree graphs, supplemented by one loop
corrections, give such a good description of many processes in QCD even at rather modest
energies where $\alpha_s$ is relatively large. Both higher order terms and
explicit non-perturbative contributions typically contribute less than a per cent; only
when one is sensitive to infrared, bound state or threshold problems is it expected that
the perturbative description becomes inappropriate.

                Finally we should draw attention to the enormous size of the
coefficients $r_N$ in Table 1; evcntually these lead to contributions that exceed the
leading term which, if taken seriously, would invalidate the predictions of asymptotic
freedom as is clear from Fig. 1.  The technique suggested here says that, in the spirit of
asymptotic series, all contributions from diagrams of $O(N_o)$ and higher should be ignored.
It could therefore be argued that, just as in the series (\ref {twelve}) where an
individual term such as $100!(0.1)^{100}$, for example, should be thought of as irrelevant
and meaningless as far as contributing to the "sum" of the series is concerned, so in (\ref
{fourteen})  8th order graphs, for example, are likewise meaningless and irrelevant (at
least until one gets to sufficiently high energies).

\begin{table}[t]
\caption{The coefficients $r_N$ occurring in the expansion of $R$, Eq. (\ref{fourteen}).
The first four are exact while the remaining are estimates from Eq. (\ref{fifteen}). Also
shown are estimates of $R_N$, the corresponding deviation from the exact result of the sum
of the first $N$ terms of the expansion (calculated using $\alpha_s=0.15$). Note that
although the $r_N$ rapidly diverge,
$R_N$ goes through a minimum near $N\sim 5$ where the error is only $6.85\times 10^{-4}$.}

\vspace{0.4cm}
\begin{center}
\begin{tabular}{|c|c|c|}  
%\multicolumn{3}{|@{\hspace{-2cm}}c|}{\sc \quad\quad\quad\quad\quad PLANT MASS}& & 
%\multicolumn{2}{c|}{\sc  BRANCH RADIUS} \\
%\cline{1-3}
%\cline{1-3}
\hline
{$N$} & {$r_{N+1}$} & {$R_N\equiv |r_{N+1}(\alpha_s/\pi)^{N+1}|$}  \\
\hline
% & & & \multicolumn{2} {c||}{\sc  E} \\
%PREDICTED &  &  &  & PREDICTED \quad OBSERVED\\
\hline
$-1$ &  $1$ & $1$ \\
\hline
$0$ & $1$ & $4.77\times 10^{-2}$   \\
\hline
$1$ &  $1.41$ & $3.21\times 10^{-3}$ \\
\hline
$2$ &  $12.8$ & $1.39\times 10^{-3}$ \\
\hline
$3$ &  $163.4$ & $8.49\times 10^{-4}$ \\
\hline
$4$ &  $2.7\times 10^{3}$ & $6.71\times 10^{-4}$ \\
\hline
$5$ &  $5.78\times 10^{4}$ & $6.85\times 10^{-4}$ \\
\hline
$6$ &  $1.53\times 10^{6}$ & $8.63\times 10^{-4}$ \\
\hline
$7$ &  $4.78\times 10^{6}$ & $1.29\times 10^{-3}$ \\
\hline
$8$ &  $1.74\times 10^{9}$ & $2.25\times 10^{-3}$ \\
\hline
$9$ &  $7.23\times 10^{10}$ & $4.45\times 10^{-3}$ \\
\hline
$10$ &  $3.38\times 10^{12}$ & $9.93\times 10^{-3}$ \\
\hline
$11$ &  $1.75\times 10^{14}$ & $2.45\times 10^{-2}$ \\
\hline
$12$ &  $9.97\times 10^{15}$ & $6.68\times 10^{-2}$ \\
\hline
$13$ &  $6.19\times 10^{16}$ & $1.98\times 10^{-1}$ \\
\hline
$14$ &  $4.17\times 10^{18}$ & $6.38\times 10^{-1}$ \\
\hline
$15$ &  $3.02\times 10^{21}$ & $2.2$ \\
\hline

\end{tabular}
\end{center}
\end{table}


\begin{thebibliography}{99}
\bibitem{foot1} When $L$ lies along the negative real axis and Im$F(z)\geq 0$ (as for a
forward scattering amplitude, for example), a rigorous bound, $R_N(z)\leq a_{N+1}z^{N+1}$,
can be derived.
\bibitem{foot2} If $G(u)$ has a singularity at $u=1$, then Eq. (\ref{ten}) suggests that
there is an explicit non-perturbative contribution to $F(z)$ which behaves like $e^{-1/z}$.
This mimics the situation in QCD where there such explicit contributions ($\sim
e^{8\pi^2/g^2}$) arising from instantons and the like.
\end{thebibliography}
\end{document}